\documentstyle[12pt]{article}
\begin{document}
\vskip 0.1in
\centerline{\Large\bf Spherically Symmetric Solutions and} 
\centerline{\Large\bf Dark Matter on the Brane}
\vskip .7in
\centerline{Dan N. Vollick}
\centerline{Department of Physics}
\centerline{Okanagan University College}
\centerline{3333 College Way}
\centerline{Kelowna, B.C.}
\centerline{V1V 1V7}
\vskip .9in
\centerline{\bf\large Abstract}
\vskip 0.5in
It has recently been suggested that our universe is a three-brane
embedded in a higher dimensional spacetime. In this paper I examine
static, spherically symmetric solutions that satisfy the effective
Einstein field equations on a brane embedded in a five dimensional
spacetime. The field equations involve a term depending on the
five dimensional Weyl tensor, so that the solutions will not be
Schwarzschild in general. This Weyl term is traceless so that
any solution of $^{(4)}R=0$ is a possible four dimensional spacetime.
Different solutions correspond to different five dimensional spacetimes
and to different induced energy-momentum tensors on the brane. One
interesting possibility is that the Weyl term
could be responsible for the observed dark matter in the universe.
\newpage
\section*{Introduction}
It has recently been suggested that some of the extra dimensions required by
string theory may be ``large'' \cite{Ar1,An1} or even infinite \cite{Ra1}.
In the scenario proposed in \cite{Ar1,An1} the spacetime is $M^{(4)}\times K$,
where $M^{(4)}$ is four dimensional Minkowski space and $K$ is a compact
manifold. The size of the extra dimensions must be $\stackrel{<}{\sim}$
$5\times 10^{-5}$ mm to be consistent with observations \cite{Cu1,Ha1}.
In the Randall and Sundrum model \cite{Ra1} our
three-brane is a domain wall separating two semi-infinite
anti-de Sitter regions. In both scenarios the standard model fields are
confined to the brane and gravity propagates in the bulk.
  
The Einstein field equations on the brane were derived by Shiromizu,
Maeda, and Sasaki \cite{Sh1}. The effective four dimensional
energy-momentum tensor contains terms involving stresses on the
brane and a term involving the five dimensional
Weyl tensor. The latter term carries information about the spacetime off the 
brane. Since the Weyl term is traceless any four dimensional spacetime
with $^{(4)}R=0$ gives rise to a three-brane world, without surface
stresses, embedded in a five dimensional spacetime. In a recent paper
Dadhich, Maartens, Papadopoulos, and Rezania \cite{Da1} examined the
Reissner-Nordstrom solution as an exact black hole solution without 
electric charge. The $g_{tt}$ component of the metric 
\begin{equation}
g_{tt}=1-\frac{2m}{r}+\frac{Q}{r^2}
\end{equation}
contains the term $Q/r^2$, which does not arise from electric charge,
but instead arises from projecting the gravitational field in
the bulk onto the brane. In general relativity $Q=q^2\geq 0$, where $q$
is the black hole charge. Here $Q$ may be zero, positive or negative.
It is important to note that there is an induced energy-momentum tensor
on the brane if $Q\neq 0$.
  
In this paper I examine the static, spherically symmetric solutions that
satisfy $^{(4)}R=0$. All of these solutions can be interpreted as
spherically symmetric solutions on the brane with the Weyl term acting as the
four dimensional energy-momentum tensor. Different solutions correspond
to different five dimensional spacetimes and different Weyl terms.
One interesting possibility that is discussed involves using the Weyl
term to account for the observed dark matter in the universe.
   
\section*{Spherically Symmetric Solutions}
The metric on the brane will be taken to be
\begin{equation}
ds^2=-B(r)dt^2+A(r)dr^2+r^2(d\theta^2+\sin^2\theta d\phi^2)
\end{equation}
and the equation $^{(4)}R=0$ gives
\begin{equation}
\frac{B^{''}}{AB}-\frac{B^{'}}{2AB}\left(\frac{A^{'}}{A}+\frac{B^{'}}
{B}\right)+\frac{2}{rA}\left(\frac{B^{'}}{B}-\frac{A^{'}}{A}\right)
+\frac{2}{Ar^2}-\frac{2}{r^2}=0 .
\label{Ricci}
\end{equation}
Of course, since we have one equation for two variables there is not
a unique asymptotically flat solution. For a given $B(r)$ the resulting
differential equation for $A(r)$ is
\begin{equation}
f(r)\frac{A^{'}}{A}+\frac{2A}{r^2}+g(r)=0
\end{equation}
where
\begin{equation}
f(r)=\frac{B^{'}}{2B}+\frac{2}{r}
\end{equation}
and 
\begin{equation}
g(r)=-\frac{B^{''}}{B}+\frac{(B^{'})^2}{2B^2}-\frac{2B^{'}}{rB}-\frac{2}{r^2}
\end{equation}
which can be integrated (in principle) to find $A(r)$. Here I will consider
a few ``types" of solutions to (\ref{Ricci}).
  
First consider solutions with $A(r)=B(r)^{-1}$. The resulting equation
is
\begin{equation}
r^2B^{''}+4rB^{'}+2(B-1)=0 .
\end{equation}
This is an Euler equation in $(B-1)$ and the general solution is
\begin{equation}
A(r)^{-1}=B(r)=1+\frac{\alpha}{r}+\frac{\beta}{r^2}
\end{equation}
where $\alpha$ and $\beta$ are constants. Thus the Reissner-Nordstrom
solution is the most general solution with $A(r)^{-1}=B(r)$.
   
Next consider spacetimes in which $B(r)$ takes the Schwarzschild form
\begin{equation}
B(r)=1-\frac{2m}{r} .
\end{equation}
The general solution for $A(r)$ is
\begin{equation}
A(r)=\left(1-\frac{2m}{r}\right)^{-1}\left[\frac{3m-2r}{\lambda-2r}
\right]
\end{equation}
where $\lambda$ is a constant. Note that Schwarzschild spacetime is
recovered if $\lambda=3m$.
  
Unless $A(r)$ and $B(r)$ correspond to the Schwarzschild solution
the brane will have a non-zero effective energy-momentum tensor
given by \cite{Sh1}
\begin{equation}
^{(4)}T_{\mu\nu}=\frac{1}{8\pi}\;^{(5)}C^{\alpha}_{\beta\rho\sigma}\eta_{\alpha}
\eta^{\rho}h_{\mu}^{\beta}h_{\nu}^{\sigma}
\end{equation}
where $^{(5)}C^{\alpha}_{\beta\rho\sigma}$ is the five dimensional Weyl
tensor evaluated on either side of the brane, $\eta_{\alpha}$ is normal to
the brane, and $h_{\mu\nu}=g_{\mu\nu}-\eta_{\mu}\eta_{\nu}$ is the
induced metric on the brane. The five dimensional spacetime off the brane
has been examined perturbatively by Sasaki, Shiromizu, and Maeda
\cite{Sa1}, but is not known in general. It is therefore more useful
to use
\begin{equation}
^{(4)}T_{\mu\nu}=-\frac{1}{8\pi}\;^{(4)}G_{\mu\nu} .
\end{equation}
It is also interesting to note that $^{(4)}T_{\mu\nu}$ does not have to 
satisfy the weak energy condition \cite{Da1,Vo1} and the spacetime may
therefore contain negative energy densities.
   
Finally consider the weak field limit with
\begin{equation}
R\simeq B^{''}+\frac{2}{r}(B^{'}-A^{'})+\frac{2}{r^2}(A^{-1}-1).
\end{equation}
Now let $A=1+a$ and $B=1+b$ where $|a|,|b|<<1$. The field equation
\begin{equation}
b^{''}+\frac{2}{r}(b^{'}-a^{'})-\frac{2}{r^2}a=0
\end{equation}
can be integrated to give
\begin{equation}
a(r)=\frac{c_1}{r}+\frac{1}{2}rb^{'}(r) ,
\label{weak}
\end{equation}
where $c_1$ is a constant.
   
One interesting possibility is that deviations 
from Schwarzschild geometry, produced by the Weyl term, could be responsible
for the observed dark matter in the universe. Since galactic rotation
curves become flat at large distances the Newtonian potential must
contain a term proportional to $\ln r$. This implies that $b(r)$ has
the form
\begin{equation}
b(r)=-\frac{2m}{r}+\alpha\ln r
\end{equation}
where $\alpha$ is a positive constant. To force $b(r)\rightarrow 0$ as 
$r\rightarrow 0$
we could replace $\ln r$ by $f(r)\ln r$ where $f(r)\simeq 1$ on galactic
scales and $f(r)\ln r\rightarrow 0$ as $r\rightarrow
\infty$. From (\ref{weak})
\begin{equation}
a(r)=\frac{c+m}{r}+\frac{1}{2}\alpha .
\end{equation}
The four dimensional energy-momentum tensor associated with the above
solution is
\begin{equation}
T_{tt}=\frac{\alpha}{16\pi r^2}
\end{equation}
\begin{equation}
T_{rr}=\frac{\alpha}{16\pi r^2}+\frac{m-c}{8\pi r^3}
\end{equation}
\begin{equation}
T_{\theta\theta}=\frac{c-m}{16\pi r}
\end{equation}
and $T_{\phi\phi}=\sin^2\theta T_{\theta\theta}$. 
Note that the energy density measured by static observers is positive if
$\alpha>0$.
\section*{Conclusion}
In this paper I examined static spherically symmetric solutions of the
equation $^{(4)}R=0$. This equation arises in ``brane-world'' models
that have zero surface stresses and a traceless source involving the
Weyl tensor. Different solutions correspond to different five dimensional
spacetimes and to different induced energy-momentum tensors on the
brane. One interesting possibility that was examined involved using the
Weyl term to account for the observed dark matter in the universe.

\end{document}